\documentclass[aps,twocolumn,superscriptaddress,showpacs,prd]{revtex4}
\usepackage{graphics}
\begin{document}
\def\Universita{Universit\`a}
\title[Scharnhorst...]{Scharnhorst effect at oblique incidence}
\author{Stefano Liberati}
\email[]{liberati@sissa.it}
\homepage[http://www.sissa.it/\~{}liberati]{}
\altaffiliation{Present address: University of Maryland, College
Park, MD 20742-4111, USA. E-mail:{\sf liberati@physics.umd.edu}}
\affiliation{International School for Advanced Studies, Via
Beirut 2-4, 34014 Trieste, Italy}
\author{Sebastiano Sonego}
\email[]{sebastiano.sonego@uniud.it}
\affiliation{{\Universita} di Udine, Via delle Scienze 208, 33100
  Udine, Italy}
\author{Matt Visser}
\email[]{visser@kiwi.wustl.edu}
\homepage[http://www.physics.wustl.edu/\~{}visser]{}
\affiliation{Physics Department, Washington University, Saint
 Louis MO 63130-4899, USA}
\date{16 October 2000; revised 13 December 2000; \LaTeX-ed \today}
\begin{abstract}
  We consider the Scharnhorst effect (anomalous photon propagation in
  the Casimir vacuum) at oblique incidence, calculating both photon
  speed and polarization states as functions of angle. The analysis is
  performed in the framework of nonlinear electrodynamics and we show
  that many features of the situation can be extracted solely on the
  basis of symmetry considerations.  Although birefringence is common
  in nonlinear electrodynamics it is not universal; in particular we
  verify that the Casimir vacuum is not birefringent at any incidence
  angle. On the other hand, group velocity is typically not equal to
  phase velocity, though the distinction vanishes for special
  directions or if one is only working to second order in the fine
  structure constant. We obtain an ``effective metric'' that is subtly
  different from previous results. The disagreement is due to the way
  that ``polarization sums'' are implemented in the extant literature,
  and we demonstrate that a fully consistent polarization sum must be
  implemented via a bootstrap procedure using the effective metric one
  is attempting to define. Furthermore, in the case of birefringence,
  we show that the polarization sum technique is intrinsically an
  approximation.
\end{abstract}
\maketitle
\def\Box{\kern0.5pt{\lower0.1pt\vbox{\hrule height.5pt width 6.8pt
    \hbox{\vrule width.5pt height6pt \kern6pt \vrule width.3pt}
    \hrule height.3pt width 6.8pt} }\kern1.5pt}
\def\half{{1\over2}}
\def\im{{\mathrm i}}
\def\a{{\mathbf a}}
\def\d{{\mathrm d}}
\def\g{{\mathrm g}}
\def\m{{\mathrm m}}
\def\x{{\mathbf x}}
\def\y{{\mathbf y}}
\def\z{{\mathbf z}}
\def\k{{\mathbf k}}
\def\v{{\mathbf v}}
\def\E{{\mathbf E}}
\def\B{{\mathbf B}}
\def\*F{{}^{\star}\/F}
\def\cs{c_{\mathrm s}}
\def\ce{c_{\mathrm{eff}}}
\def\L{{\cal L}}
\def\I{{\cal I}}
\def\H{{\cal H}}
\def\ie{{\emph{i.e.}}}
\def\eg{{\emph{e.g.}}}
\def\Schwinger{{\mathrm{Schwinger}}}
\def\Casimir{{\mathrm{Casimir}}}
\def\background{{\mathrm{background}}}
\def\electron{{\mathrm{e}}}
\def\phase{{\mathrm{phase}}}
\def\group{{\mathrm{group}}}
\def\signal{{\mathrm{signal}}}
\def\effective{{\mathrm{effective}}}
\def\photon{{\mathrm{photon}}}
\def\HRULE{{\bigskip\hrule\bigskip}}
\section{Introduction}
In 1990 Scharnhorst demonstrated that the propagation of light in
the Casimir vacuum~\cite{Casimir,Mostepanenko,Casimir-report} is
characterized by an anomalous speed~\cite{Sch90}. In fact photons
propagating perpendicular to the plates travel at a speed
$c_{\bot}$ which slightly exceeds the usual speed of light $c$ in
the Minkowski vacuum~\cite{Sch90,Bar90,SchBar93,Sch98}. The
propagation of photons parallel to the plates instead occurs at
the usual speed $c_{\parallel}=c$.

Unfortunately this anomalous propagation is far too small to be
experimentally detectable; the relative modifications to the
speed of light are of order $10^{-2}\alpha^2/(m_\electron
\,a)^4$, where $\alpha$ is the fine structure constant, $a$ is
the distance between the plates, and $m_\electron$ is the
electron mass. [We work in units such that $\hbar=c=1$. Greek and
Latin indices run from 0 to 3 and from 1 to 3, respectively.
$\eta_{\mu\nu}= \mbox{diag} \left(-1,+1,+1,+1\right)$ denotes the
Minkowski metric.] It is nevertheless an important point of
principle that a quantum vacuum which is polarized by an external
constraint can behave as a dispersive medium with refractive
index $n(\omega)$ which remains less than unity to arbitrarily
high frequency. Similar effects have been discovered in the case
of quantum vacuum polarization in gravitational
fields~\cite{DH80,DS94,DS96,Shore96}, and several other
situations~\cite{Latorre}.

It is to be emphasized that the Scharnhorst effect, albeit small,
is of fundamental theoretical importance.  Though the present
calculations are carried out in the ``soft photon'' approximation
(wavelengths much larger than an electron Compton wavelength)
there is an argument based on dispersion relations which strongly
suggests the actual signal speed is also modified (indeed,
increased)~\cite{SchBar93,Sch98} --- the physics here is very
different from that of the resonance-induced ``apparent''
superluminal velocities currently of experimental
interest~\cite{apparent-superluminal,no-thing}; here we are
dealing with quantum polarization induced ``true'' superluminal
velocities, albeit well outside the realm of present day
experimental technique.

A common feature of the gravitational analogs of the Scharnhorst
effect is the presence of birefringence: Photons with different
polarizations propagate at different speeds.  Similarly in generic
situations nonlinear electrodynamics often leads to
birefringence~\cite{Adler,Novello,Novello-bis}. Nevertheless the
occurrence of birefringence in nonlinear electrodynamics is not
universal. Indeed, in the Casimir vacuum between parallel conducting
plates the possibility of birefringence is tightly constrained: There
is a residual (2+1) Lorentz invariance which prevents birefringence
for photons that propagate parallel to the plates, and similarly there
is an $O(2)$ rotational invariance that prevents birefringence for
photons that propagate perpendicular to the plates. It is only for
photons that propagate obliquely to the plates that there is even any
possibility of birefringence.  We shall be interested in a complete
analysis of both propagation speed and polarization states as a
function of angle, and as a side effect report the (melancholy)
conclusion that birefringence is completely absent in the Casimir
vacuum.  As a consequence, an effective metric description is
sufficient for completely specifying photon propagation in the Casimir
vacuum---though it will in general not be sufficient for other more
general vacuum states. The effective metric we find is subtly at
variance with previous results obtained by performing a ``polarization
sum''~\cite{Dittrich-Gies,DGbook}.  The disagreement is due to the way
that ``polarization sums'' are implemented in the extant literature,
where they are defined in terms of the flat Minkowski metric.  We show
that in the absence of birefringence a fully consistent polarization
sum must be implemented in terms of a bootstrap procedure; using as
input the effective metric one is attempting to define. Furthermore,
in the presence of birefringence, we demonstrate that the polarization
sum technique is intrinsically an approximation.  Fortunately the
differences first show up at order $\alpha^4$, and in almost all
circumstances are completely negligible.

Throughout, we emphasize the use of symmetry arguments as a way of
extracting general information that is (as much as possible)
independent of the particular choice of Lagrangian. Finally, we
have a few words to say concerning the utility of effective
metric approaches in other contexts well beyond the Casimir
vacuum.

\section{General Formalism}
\label{S:general}
Anomalous photon propagation can most easily be interpreted in
terms of nonlinear electrodynamics.  After integrating out
virtual electron loops, the Maxwell Lagrangian should be replaced
(in the absence of boundaries, at the one-loop level, and
provided the distance scales defined by the field gradients and
photon wavelength are much larger than the electron Compton
wavelength) by Schwinger's effective Lagrangian~\cite{Schwinger}
\begin{equation}
\label{E:Schwinger-lagrangian} \L_\Schwinger = \L\left({\cal
F},{\cal G}\right).
\end{equation}
Here we have adopted the now common variables~\cite{Dittrich-Gies,DGbook}
\begin{eqnarray}
\label{E:xy} {\cal F} &\equiv & \frac{1}{4} F_{\mu\nu} \;
F^{\mu\nu} = \frac{1}{2}\left(\vec{B}^2-\vec{E}^2\right),
\\
{\cal G} &\equiv & \frac{1}{4} F_{\mu\nu} \; \*F^{\mu\nu} =
-\vec{E}\cdot \vec{B}.
\end{eqnarray}
The precise functional form of Schwinger's Lagrangian will not be
needed: At the end of the calculation we shall see that it is
sufficient to retain only the quartic terms beyond the Maxwell
Lagrangian (the Euler--Heisenberg Lagrangian~\cite{Euler-Heisenberg}).

Now in the Casimir geometry between parallel plates there is one
additional invariant one could consider, namely
\begin{eqnarray}
\label{E:z} {\cal H} &\equiv & n^\mu  F_{\mu\sigma} \;
F^{\sigma\nu} \; n_\nu = (\vec{E}\cdot\vec{n})^2 - (\vec{n}
\times \vec{B})^2,
\end{eqnarray}
where $n^\mu \equiv(0,0,0,1)$ is a unit vector orthogonal to the
plates. (A similar invariant obtained by interchanging $\vec{E}$
and $\vec{B}$ is actually a redundant linear combination of the
invariants $\cal F$ and $\cal H$.) In addition the coefficients
in the effective Lagrangian can now explicitly depend on the $z$
coordinate. When it comes to practical calculations all these
effects are suppressed by additional factors of
$\hbar/(m_\electron c a)$, but in the interests of generality we
retain them for the time being and simply write:
\begin{equation}
\label{E:effective-lagrangian} \L_\effective =
\L\left(F_{\mu\nu}(x),x\right).
\end{equation}
That is, the considerations of the following section are not
limited to the Casimir parallel plate geometry but apply whenever
the soft photon approximation (and the ancillary linearization
procedure and restricted eikonal approximation) make sense.

\subsection{Equations of motion}

The complete equations of motion for nonlinear electrodynamics
consist of the Bianchi identity,
\begin{equation}
\label{E:bianchi} F_{[\mu\nu,\lambda]}=0,
\end{equation}
plus the dynamical equation
\begin{equation}
\label{E:eom0}
\partial_{\nu} \left( {\partial\L\over\partial F_{\mu\nu} }\right) =0.
\end{equation}
We now adopt a linearization procedure: Split the electromagnetic
field into a background plus a propagating photon
\begin{equation}
\label{E:linearize} F_{\mu\nu}= F_{\mu\nu}^{\,\background} +
f_{\mu\nu}^{\,\photon}.
\end{equation}
Then assuming the background satisfies the equations of motion, and
retaining only linear terms in the propagating photon, we have
\begin{equation}
\label{E:bianchi2} f^{\,\photon}_{[\mu\nu,\lambda]}=0,
\end{equation}
and
\begin{equation}
\label{E:eom1}
\partial_{\nu}
\left( \left. {\partial^2\L\over\partial F_{\mu\nu} \; \partial
F_{\alpha\beta} } \right|_\background \;
f^{\,\photon}_{\alpha\beta} \right) =0.
\end{equation}
On defining
\begin{equation}
\label{E:define-omega} \Omega^{\mu\nu\alpha\beta}= \left.
{\partial^2\L\over\partial F_{\mu\nu} \; \partial F_{\alpha\beta}}
\right|_\background,
\end{equation}
equation (\ref{E:eom1}) can be rewritten in the somewhat more
compact form
\begin{equation}
\label{E:eom2}  \partial_\alpha \left(\Omega^{\mu\alpha\nu\beta}\;
f^{\,\photon}_{\nu\beta} \right)=0.
\end{equation}
Note that the tensor $\Omega^{\mu\nu\alpha\beta}$ is symmetric
with respect to exchange of the pairs of indices $\mu\nu$ and
$\alpha\beta$, and antisymmetric with respect to exchange of
indices within each pair.

We now apply a restricted form of the eikonal approximation by
introducing a slowly varying amplitude $f_{\mu\nu}$ and a rapidly
varying phase $\phi$:
\begin{equation}
\label{E:eikonal} f^{\,\photon}_{\mu\nu} = f_{\mu\nu}\; {\rm
e}^{\im\phi}.
\end{equation}
The wave vector (actually a one-form, but we shall loosely refer
to both vectors and one-forms as ``vectors'' in the text, because
of the mapping provided by $\eta_{\mu\nu}$ and its inverse
$\eta^{\mu\nu}$) is then defined as $k_{\mu}=\partial_{\mu}\phi$.
This approximation is similar to, but not quite identical with,
the usual eikonal approximation. This is because one assumes that
$\phi$ varies on scales much smaller than those of the
background, while, on the other hand, use of the Lagrangian
(\ref{E:effective-lagrangian}) also implies that the components
of $k$ are much smaller than the values fixed by the electron
mass (soft-photon regime).  Under these hypotheses,
\begin{equation}
\label{E:eom3}
\Omega^{\mu\alpha\nu\beta}\;k_{\alpha} \;
f_{\nu\beta}=0.
\end{equation}
But the background field is itself subject to quantum fluctuations,
and to take this into account the coefficients of this equation are
identified with the expectation value of the corresponding quantum
operators in the background state $|\psi\rangle$:
\begin{equation}
\label{E:eom-master}
\langle\psi |
\Omega^{\mu\alpha\nu\beta}|\psi\rangle \;k_{\alpha} \;
f_{\nu\beta}=0.
\end{equation}
In taking this expectation value we are using the fact that the
fluctuations in the background fields are determined by the
geometry (in the specific case of the Casimir geometry, by the
distance between the plates), so that in the spirit of the
restricted eikonal approximation there is a separation of scales
between the background fluctuations and the propagating photon.

The Bianchi identity (\ref{E:bianchi2}) constrains $f_{\mu\nu}$ to
be of the form
\begin{equation}
\label{E:fgp}
f_{\mu\nu} = k_{\mu} \;a_{\nu}-k_{\nu}\; a_{\mu}\,,
\end{equation}
where we have introduced the gauge potential $a$ for the
propagating field. Inserting (\ref{E:fgp}) into (\ref{E:eom-master})
we find
\begin{equation}
\label{E:eom-master2}
\langle\psi |
\Omega^{\mu\alpha\nu\beta}|\psi\rangle \;k_{\alpha} \;
k_\beta\;a_\nu=0.
\end{equation}

In general, the tensorial quantity $\langle\psi
|\Omega^{\mu\alpha\nu\beta}| \psi\rangle$ can be decomposed into
an isotropic part plus anisotropic contributions, that we group
together into a term $\Delta^{\mu\alpha\nu\beta}$ with the same
symmetries as $\Omega^{\mu\alpha\nu\beta}$:
\begin{equation}
\label{E:define-delta}
\langle \psi
|\Omega^{\mu\alpha\nu\beta}|\psi\rangle = d_1 \, \big(
\eta^{\mu\nu}\; \eta^{\alpha\beta}- \eta^{\mu\beta} \;
\eta^{\alpha\nu} \big)+\Delta^{\mu\alpha\nu\beta},
\end{equation}
where $d_1$ is a function that can in principle be computed
directly from the effective Lagrangian.

\subsection{Dispersion relations}
\label{SS:dispersion-relations}

Equation (\ref{E:eom-master2}) represents a condition for $a$ as a
function of $k$ --- it constrains $a$ to be an eigenvector, with
zero eigenvalue, of the $k$-dependent matrix
\begin{equation}
\label{E:matrix} A^{\mu\nu}(k)= \langle\psi |
\Omega^{\mu\alpha\nu\beta}|\psi\rangle \; k_\alpha \; k_\beta.
\end{equation}
Any non-zero solution corresponds to a physically possible field
polarization, that can be identified by a unit polarization vector
$\epsilon$ (provided $a$ is not a null vector --- a possibility
that can always be avoided by a suitable gauge choice).

A necessary and sufficient condition for the eigenvalue problem
$A^{\mu\nu}\,a_\nu=0$ to have non-zero solutions is
$\mbox{det}\left(A^{\mu\nu} \right)=0$; however, this gives us no
information at all. Indeed, any $a$ parallel to $k$ is always a
non-zero solution, so the condition $\mbox{det}\left(A^{\mu\nu}
\right)=0$ is actually an identity. On the other hand,
$a\parallel k$ is merely an unphysical gauge mode that
corresponds to $f_{\mu\nu}=0$ by (\ref{E:fgp}), so we need to find
other, physically meaningful, solutions of the eigenvalue
problem. To this end, we exploit gauge invariance under $a \to a
+\lambda\, k$ and fix a gauge, thus removing the spurious modes.
It is particularly convenient to adopt the temporal gauge
$a_0=0$. Then we can define a polarization vector
$\epsilon_\mu\equiv a_\mu/\left(a_\nu a^\nu\right)^{1/2}$, and the
eigenvalue problem $A^{\mu\nu}\,\epsilon_\nu=0$ splits into the
equation
\begin{equation}
\label{E:00} A^{0i}\,\epsilon_i=0,
\end{equation}
plus the reduced eigenvalue problem
\begin{equation}
\label{E:reduced} A^{ij}\,\epsilon_j=0\,.
\end{equation}
The latter admits a nontrivial solution only if
\begin{equation}
\label{E:detA=0} \mbox{det}\left(A^{ij}\right)=0.
\end{equation}
The condition (\ref{E:detA=0}) plays the same role as the Fresnel
equation in crystal optics~\cite{landau} --- it is a scalar
equation for $k$ and thus gives the dispersion relation for light
propagating in our ``medium''.

An explicit calculation (see Appendix~\ref{app:fresnel}) gives
\begin{equation}
  \label{eq:poli}
  \mbox{det}\left( A^{ij}\right)=\omega^2 \; {\cal P}_{4}(k),
\end{equation}
where $\omega\equiv -k_0$ and ${\cal P}_{4}(k)$ is a homogeneous
fourth-order polynomial in the variables $\omega$ and $k_{i}$.
This means that in the most general case there are four dispersion
relations, corresponding to the four roots of the equation
\begin{equation}
{\cal P}_{4}(k)=0.
\end{equation}
But, if $k=(-\omega,\vec k)$ is a root then so is
$-k=(\omega,-\vec k)$; thus, by CPT invariance, only two of these
dispersion relations are physically distinct.

Different polarization states are represented by linearly
independent solutions of the eigenvalue problem
(\ref{E:reduced}), under the condition (\ref{E:detA=0}).
[Obviously, (\ref{E:00}) cannot be independent of
(\ref{E:reduced}), since we know that
$\mbox{det}\left(A^{\mu\nu}\right) \equiv 0$.]  Thus, the space of
polarizations is at most two-dimensional.  Since (\ref{E:detA=0})
gives rise to two dispersion relations, the polarization states
actually satisfy two (in general, different) eigenvalue equations,
\begin{equation}
\label{E:nonsing}
\overline{A}_{(r)}^{\,\mu\nu}\;\epsilon^{\,(r)}_{\;\nu}=0\,,
\end{equation}
where $r=1,2$ labels the dispersion relations and
$\overline{A}_{(r)}^{\,\mu\nu}$ is obtained from $A^{\mu\nu}$ by
imposing the corresponding condition on $k$.  Thus, in the general
case, modes of the field with different polarizations have
different dispersion relations, hence propagate in different
ways; this leads to the phenomenon of
birefringence~\cite{Adler,Novello,Novello-bis}.

In some special cases (in particular, this behaviour is quite
common in nonlinear electrodynamics) the polynomial ${\cal
P}_{4}(k)$ factorizes into two quadratic forms,
\begin{equation}
  \label{eq:factor}
{\cal P}_{4}(k)= \bigl(\gamma_{(1)}^{\,\mu\nu} \; k_\mu \, k_\nu
\bigr)  \; \bigl(\gamma_{(2)}^{\,\alpha\beta} \; k_\alpha \,
k_\beta \bigr),
\end{equation}
in which case we obtain two second-order dispersion relations:
\begin{equation}
  \label{eq:disprel}
  \gamma_{(1)}^{\,\mu\nu} k_\mu k_\nu=0 \quad
   \mbox{and} \quad
  \gamma_{(2)}^{\,\mu\nu} k_\mu k_\nu=0.
\end{equation}

\subsection{Effective metric}
\label{SS:effmet}

We now want to consider the special situation in which not only
does (\ref{eq:factor}) hold, but also $\gamma_{(1)}^{\,\mu\nu}=
\gamma_{(2)}^{\,\mu\nu}$. (That is, the fourth-order polynomial
${\cal P}_{4}(k)$ is a perfect square.) In this case one ends up
with a single quadratic dispersion relation of the familiar form
\begin{equation}
 \label{E:eomfc}
 \gamma^{\,\mu\nu}\,k_{\mu}\,k_{\nu}=0,
\end{equation}
where $\gamma^{\,\mu\nu}$ is some symmetric tensor.  It should be
clear from our previous discussion that a necessary condition for
this to happen is the absence of birefringence. Remarkably, we
see that the wave vector is now null with respect to a (unique)
``effective inverse metric'' $\gamma^{\,\mu\nu}$. Therefore, the
propagation of light can be described in terms of an effective
geometry, defined by the metric tensor ${\rm g}_{\mu\nu}$
obtained by inverting $\gamma^{\,\mu\nu}$, such that
$\gamma^{\,\mu\nu}{\rm g}_{\nu\rho}=\delta^{\,\mu}{}_\rho$.

(Warning: Even when an effective metric is defined, we always
raise and lower indices using the flat Minkowski metric
$\eta_{\mu\nu}$.  In particular, note that ${\rm g}_{\mu\nu}\neq
\eta_{\mu\rho}\,\eta_{\nu\sigma}\gamma^{\,\rho\sigma}$; this
justifies our use of different symbols, $\rm g$ and $\gamma$, for
the effective metric and its inverse.)

This situation implies that
$\langle\psi|\Omega^{\mu\alpha\nu\beta}|\psi\rangle$ must be
algebraically constructible solely in terms of
$\gamma^{\,\mu\nu}$. In view of the symmetries of
$\Omega^{\mu\alpha\nu\beta}$ we know, without need for detailed
calculation, that it must be of the form
\begin{equation}
 \label{E:Omega2}
 \langle\psi|\Omega^{\mu\alpha\nu\beta}|\psi\rangle =\Lambda
 \bigl(\gamma^{\,\mu\nu} \; \gamma^{\,\alpha\beta}
 - \gamma^{\,\mu\beta} \; \gamma^{\,\nu\alpha}\bigr)
\end{equation}
for some function $\Lambda$.

Conversely, if $\Omega^{\mu\alpha\nu\beta}$ is of the form
(\ref{E:Omega2}), then the matrix (\ref{E:matrix}) is
\begin{equation}
A^{\mu\nu}=\Lambda\,\left[\gamma^{\,\mu\nu}
\bigl(\gamma^{\,\alpha\beta}\,k_\alpha\,k_\beta\bigr)-
\bigl(\gamma^{\,\mu\alpha}\,k_\alpha\bigr)
\bigl(\gamma^{\,\nu\beta}\,k_\beta\bigr)\right],
\end{equation}
and the photon propagation equation $A^{\mu\nu}\,a_\nu=0$ becomes
\begin{equation}
\label{E:newpropageq}
\bigl(\gamma^{\,\alpha\beta}\,k_\alpha\,k_\beta\bigr)
\gamma^{\,\mu\nu}\,a_\nu-
\bigl(\gamma^{\,\alpha\beta}\,a_\alpha\,k_\beta\bigr)
\gamma^{\,\mu\nu}\,k_\nu=0.
\end{equation}
This equation is obviously satisfied by the uninteresting gauge
modes $a\parallel k$, with no constraints on $k$. Solutions
corresponding to a non-vanishing $f_{\mu\nu}$ exist only if the
coefficient of $\gamma^{\,\mu\nu}\,a_\nu$ is zero, \ie, if
(\ref{E:eomfc}) holds. Thus, the two polarization states
propagate with the same dispersion relation (\ref{E:eomfc}), and
there is no birefringence.

Substituting (\ref{E:eomfc}) back into the propagation equation
(\ref{E:newpropageq}) we find another relationship typical of this
case,
\begin{equation}
\label{E:gkepsilon} \gamma^{\,\mu\nu}\,k_\mu\,a_\nu=0.
\end{equation}
Formally, the above equation looks like a gauge condition. This
might seem puzzling, because nowhere in the present subsection
have we fixed a gauge. In fact, (\ref{E:gkepsilon}) is a
consequence of the dynamical equation $A^{\mu\nu}\,a_\nu=0$, when
the ``on-shell'' condition (\ref{E:eomfc}) is satisfied, and it
does {\em not\/} imply any gauge fixing.

Finally, it is interesting to notice that there is now a
self-consistency or ``bootstrap'' condition,
\begin{equation}
 \label{E:bootstrap2}
3 \,\Lambda\, \gamma^{\,\mu\nu} =
\langle\psi|\Omega^{\mu\alpha\nu\beta}|\psi\rangle\; {\rm
g}_{\alpha\beta}.
\end{equation}
This can be re-stated directly in terms of the fundamental
coefficients as
\begin{eqnarray}
\label{E:bootstrap3} 3\,\Lambda\, \gamma^{\,\mu\nu} &=& d_1
\;\eta^{\mu\nu} \left( \eta^{\alpha\beta} \; {\rm g}_{\alpha\beta}
\right) - d_1 \; \eta^{\mu\alpha}\; {\rm g}_{\alpha\beta} \;
\eta^{\beta\nu} \nonumber
\\
&& \qquad + \Delta^{\mu\alpha\nu\beta} \; {\rm g}_{\alpha\beta}.
\end{eqnarray}
We stress that these relations depend only on the assumed
existence of a single unique effective metric ${\rm g}_{\mu\nu}$
--- they do not make any reference to other specifics of the
quantum state.

\section{Casimir vacuum}
Let us now consider a region of empty space delimited by two
perfectly conducting parallel plates placed orthogonal to the $z$
axis at positions $z=z_0$ and $z=z_0+a$, in the quantum state
$|\psi\rangle \to |C\rangle$ corresponding to the Casimir
vacuum~\cite{Casimir,Mostepanenko,Casimir-report}.

\subsection{Symmetries and effective metric}
In the Casimir vacuum considerable information can be extracted by
using only symmetry considerations, similarly to what Bryce
DeWitt did for the stress-energy-momentum tensor in
reference~\cite{DeWitt}.  Always working in the soft photon
approximation, the presence of a preferred direction and the
symmetry of the configuration allow us to claim that the function
$d_1$ can only depend on the $z$ coordinate, and to write
$\Delta^{\mu\alpha}{}_{\nu\beta}$ in the form
\begin{eqnarray}
\label{E:MEHsym} \Delta^{\mu\alpha}{}_{\nu\beta} &=& d_2(z) \;
\big( \delta^{\mu}{}_{\nu}\;n^{\alpha}\;n_{\beta}-
\delta^{\mu}{}_{\beta}\;n^{\alpha}\;n_{\nu}
\nonumber\\
&& \quad + \delta^{\alpha}{}_{\beta}\;n^{\mu}\;n_{\nu}-
\delta^{\alpha}{}_{\nu}\;n^{\mu}\;n_{\beta} \big),
\end{eqnarray}
where $d_2(z)$ is another function.  The calculation of $d_1(z)$
and $d_2(z)$ in this case is straightforward, using identities
that follow from (\ref{E:define-delta}) and (\ref{E:MEHsym}):
\begin{equation}
\label{E:porc1} d_1(z) = \frac{1}{6}\left\langle C\left|
\Omega^{\mu\nu}{}_{\mu\nu} - 2\,\Omega^{\mu\nu}{}_{\alpha\nu}\; n_\mu \;
n^\alpha
 \right| C\right\rangle;
\end{equation}
\begin{equation}
\label{E:porc2} d_2(z) = -\frac{1}{6}\left\langle C\left|
\Omega^{\mu\nu}{}_{\mu\nu} - 4\,\Omega^{\mu\nu}{}_{\alpha\nu}\; n_\mu \;
n^\alpha \right| C\right\rangle.
\end{equation}
On defining the function
\begin{equation}
\xi(z)\equiv {d_2(z)\over d_1(z)},
\end{equation}
and the tensor
\begin{equation}
  \label{E:metrica}
   \gamma^{\,\mu\nu}=\eta^{\mu\nu}+\xi(z)\, n^{\mu}n^{\nu},
\end{equation}
$\langle C|\Omega^{\mu\alpha\nu\beta}|C\rangle$ takes the
particularly simple form
\begin{equation}
 \label{E:Omega2c}
 \langle C|\Omega^{\mu\alpha\nu\beta}|C\rangle = d_1(z) \;
 \bigl(\gamma^{\,\mu\nu} \; \gamma^{\,\alpha\beta}
 - \gamma^{\,\mu\beta} \; \gamma^{\,\nu\alpha}\bigr).
\end{equation}
Thus, from the discussion in subsection \ref{SS:effmet}, we deduce
without further argument that there is no birefringence, and that
$\gamma^{\,\mu\nu}$ given in (\ref{E:metrica}) plays the role of
an effective inverse metric for the propagation of light in the
Casimir vacuum.  The corresponding effective metric is
\begin{equation}
\label{E:nuova!}
{\rm
g}_{\mu\nu}=\eta_{\mu\nu}-\frac{\xi(z)}{1+\xi(z)}\,n_\mu n_\nu.
\end{equation}

We stress that for this derivation of the effective metric to
make sense it is necessary that the spatial variation in $\xi(z)$
be compatible with the restricted eikonal approximation: that is
$\xi(z)$ should vary slowly on the scale of the photon
wavelength. This is certainly true for QED at lowest nontrivial
order where we shall soon see that $\xi$ is in fact independent
of $z$.

The conclusion about the absence of birefringence can also be
obtained more explicitly. Choosing the temporal gauge, and
following the steps outlined in the general analysis of subsection
\ref{SS:dispersion-relations}, equation (\ref{E:00}) becomes
\begin{equation}
\label{E:0} \gamma^{\,ij}\,k_{i}\,\epsilon_{j}=0,
\end{equation}
which also follows from (\ref{E:gkepsilon}) when $\epsilon_0=0$. A
tedious but straightforward calculation gives
\begin{equation}
\label{E:detA} \mbox{det}\left(A^{ij}\right) =-\left(1+\xi\right)
\; d_1^{\,3} \; \omega^2 \; \bigl(\gamma^{\,\mu\nu}\,
k_{\mu}\,k_{\nu} \bigr)^2.
\end{equation}
Thus the dispersion relation determined from (\ref{E:detA=0})
takes in this case the form (\ref{E:eomfc}), as expected.

We note here that, although an equation of the form
(\ref{E:eomfc}) has been already obtained by Dittrich and
Gies~\cite{Dittrich-Gies}, the expression for the effective
metric underlying their dispersion relation is subtly at variance
with our results (\ref{E:metrica}) and (\ref{E:nuova!}).  We shall
comment about the reason for this discrepancy in
Appendix~\ref{app:polsum}.

\subsection{Polarization states}
In order to explicitly find the polarization states, let us
evaluate (\ref{E:matrix}) using the expression for $\langle
C|\Omega^{\mu\alpha\nu\beta}|C\rangle$ and (\ref{E:eomfc}), and
then consider the reduced eigenvalue problem (\ref{E:reduced}).
We find that $\vec{\epsilon}$ must be a solution of the problem
$\overline{A}^{\,ij}\,\epsilon_j=0$, where
\begin{equation}
\overline{A}^{\,ij}=-\gamma^{\,il}\,k_{l}\;\gamma^{\,jm}\,k_{m}.
\end{equation}
This eigenvalue problem, however, is manifestly equivalent to the
single equation (\ref{E:0}), which is thus the only constraint
that the polarization states must satisfy. Therefore, the space of
polarization states is two-dimensional, as expected.

A basis for such a space can be easily constructed by considering
generic propagation in the $xz$ plane [not a restrictive
hypothesis, because of $O(2)$ invariance with respect to
rotations around the $z$ axis], described by the 3-vector (these
are taken to be covariant components, index down):
\begin{equation}
\label{E:wavevector} \vec{k} = \left(|\vec k|\sin\theta,0, |\vec
k|\cos\theta\right).
\end{equation}
Equation (\ref{E:0}) then becomes
\begin{equation}
\label{E:01}
\sin\theta\;\epsilon_1+\left(1+\xi\right)\,\cos\theta\;\epsilon_3=0,
\end{equation}
so two independent polarizations are:
\begin{eqnarray}
\vec{\epsilon}^{\;(1)} &=& (0,1,0);
\nonumber\\
\vec{\epsilon}^{\;(2)} &=& \frac{1}{N(\theta)}
\bigl(\left(1+\xi\right)\cos\theta,0,-\sin\theta\bigr),
\end{eqnarray}
where
\begin{equation}
N(\theta)\equiv
\left(1+2\,\xi\,\cos^2\theta+\xi^2\cos^2\theta\right)^{1/2}
\end{equation}
is a normalization coefficient.  Note in particular that
$\vec{\epsilon}^{\;(2)}$ is not perpendicular to $\vec{k}$ when
viewed in terms of the Minkowski metric $\eta^{\mu\nu}$, though
they are perpendicular when viewed in terms of the effective
metric $\gamma^{\,\mu\nu}$. Furthermore we have chosen to make
$\vec{\epsilon}^{\;(2)}$ a unit vector with respect to the
Minkowski metric, not with respect to the effective metric.

\subsection{Phase, signal, and group velocities}
If $\vec{k}$ has the form (\ref{E:wavevector}), equation
(\ref{E:eomfc}) becomes (the norms are with respect to the
Euclidean spatial metric induced by $\eta_{\mu\nu}$):
\begin{equation}
\label{E:eomcepthe2} \omega^2 = |\vec{k}|^2+ \xi\; |\vec{k}|^2 \;
\cos^{2}\theta.
\end{equation}
The phase velocity is given by
\begin{equation}
\label{E:speed} v_\phase(\theta) =\frac{\omega}{|\vec{k}|}
=\left(1+\xi\,\cos^{2}\theta\right)^{1/2},
\end{equation}
and is independent of the polarization.  This is again a
consequence of the fact that the quantity
$\gamma^{\,\mu\nu}\,k_{\mu}\,k_{\nu}$ appears squared in
$\mbox{det}\left(A^{ij} \right)$, so equation (\ref{E:detA=0})
describes a degenerate fourth-degree surface in the space of the
vectors $\vec{k}/\omega$. Hence, there is only one dispersion
relation, equation (\ref{E:eomcepthe2}), independent of the
polarization state, and only one phase speed for each value of
$|\vec{k}|$.  This confirms that birefringence does not take
place in the Casimir vacuum.

Note that if $\xi>0$ and $\theta\neq\pi/2$, we have
$v_\phase(\theta)>1$.  Since, in the limit $\omega\to +\infty$,
$v_\phase$ equals the front velocity (signal
velocity)~\cite{Sch98}, one is tempted to argue that the
propagation is superluminal for all values of $\theta$ different
from $\pi/2$.  This conclusion can also be inferred directly from
(\ref{E:eomfc}), which for $\xi>0$ implies that $k$ is timelike
(with respect to the undisturbed Minkowski metric).  Since $k$
can also be interpreted as the four-vector orthogonal to a
surface of discontinuity of the field~\cite{Novello,Novello-bis},
it follows that such a surface is spacelike (with respect to the
Minkowski metric), \ie, that electromagnetic signals travel
``faster than light''. (This phrasing is standard but
unfortunate, and is logically indefensible.  It should always be
interpreted in the sense ``faster than light would have travelled
in an undisturbed portion of normal vacuum''.)  Equivalently,
(\ref{E:nuova!}) implies that for $\xi>0$ the lightcones of the
effective metric ${\rm g}_{\mu\nu}$ are wider than those of the
Minkowski metric $\eta_{\mu\nu}$. Unfortunately, while it is
certainly true that our treatment is valid at high frequencies
(with respect to those associated to the background scales), it
nevertheless also requires $\omega\ll m_\electron$
--- the condition under which one can use the Lagrangian
(\ref{E:Schwinger-lagrangian})
---, so we have no direct information about the strict $\omega\to
+\infty$ limit.  See, however, reference~\cite{SchBar93} for an
indirect argument based on the Kramers--Kronig dispersion relation
that combined with the present calculation is sufficient to
establish superluminal propagation for the signal velocity. In
the special case $\theta=\pi/2$ one has $\eta^{\mu\nu}k_\mu k_\nu
=0$: Photons propagating parallel to the plates travel at the
standard speed of light.

The group velocity is a little tricky, it equals
\begin{equation}
v^i_{\mathrm{group}} \equiv {\partial\omega\over\partial k_i} =
\frac{1}{\omega}\left[ k^i + \xi\; (k\cdot n)\; n^i\right],
\end{equation}
so equations (\ref{E:wavevector}) and (\ref{E:speed}) give
\begin{equation}
\vec{v}_{\mathrm{group}}(\theta)=\frac{1}{v_\phase(\theta)}
\bigl(\sin\theta,0,(1+\xi)\cos\theta\bigr).
\end{equation}
In particular, the group velocity $\vec{v}_{\mathrm{group}}$ is
not parallel to the wave-vector $\vec k$, though it is always
orthogonal to the polarization vector. Taking the norm (in the
physical Minkowski metric),
\begin{equation}
\label{E:vgroup} v_{\mathrm{group}}(\theta) = {N(\theta) \over
v_\phase(\theta)} = \left[{1+(2\xi+\xi^2)\cos^2\theta\over
1+\xi\;\cos^2\theta}\right]^{1/2}.
\end{equation}
The group speed is at all angles slightly greater than (or at
worst equal to) the phase speed. (See figure 1.) Indeed
\begin{equation}
\label{E:vgroup2}
v_{\mathrm{group}}^2(\theta) =
v_\phase^2(\theta) + {\xi^2 \cos^2\theta \; \sin^2\theta
\over v_\phase^2(\theta)}.
\end{equation}
(See figure 2.) Care should be taken to realize that here
$\theta$ is the angle between the wave vector and the normal, and
is not quite the same as the direction of propagation of the wave
packet. Fortunately at both normal incidence ($\theta=0$) and
parallel propagation ($\theta=\pi/2$) the distinction between
group and phase velocities disappears. Furthermore, as we shall
see below, if we work to second order in the fine structure
constant, the difference between group and phase speed is
negligible at all angles, although a difference of order
$\alpha^2$ still remains between the directions of
$\vec{v}_{\mathrm{group}}$ and $\vec k$. The group velocity
(\ref{E:vgroup}) is greater than 1 when $\xi>0$. Thus, the group
speed is larger than 1 whenever the phase speed is, although in
general their individual values are different.

\begin{figure}[htbp]
\vbox{
\hfil
\scalebox{0.40}{\rotatebox{270}{{\includegraphics{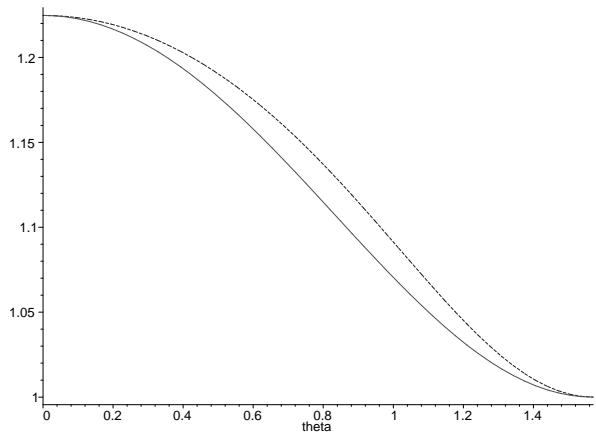}}}}
\hfil
}
\bigskip
\caption{
Group and phase speeds as a function of angle, from $\theta=0$ to
$\theta=\pi/2$ radians. (The group speed is the upper curve.) For
clarity we have greatly exaggerated the physically expected value
of $\xi$ by setting $\xi = 1/2$.
}
\label{F:speed}
\end{figure}

\begin{figure}[htbp]
\vbox{
\hfil
\scalebox{0.40}{\rotatebox{270}{{\includegraphics{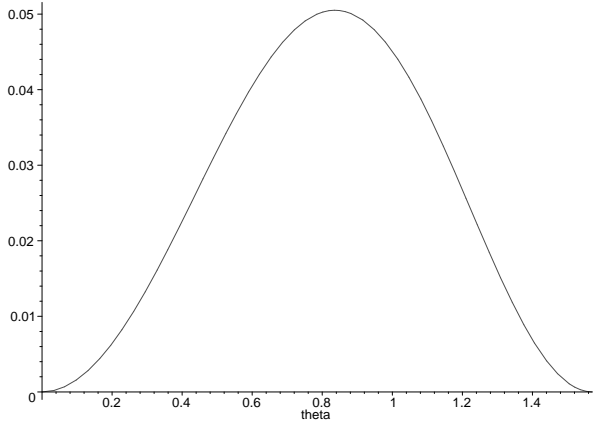}}}}
\hfil
}
\bigskip
\caption{
Difference between the squares of the group and phase speeds as a
function of angle, from $\theta=0$ to $\theta=\pi/2$ radians. For
clarity we have greatly exaggerated the physically expected value
of $\xi$ by setting $\xi = 1/2$.
}
\label{F:delta}
\end{figure}

\section{Size of the effect}
\label{S:size}
So far we have seen that, based solely on symmetry considerations, we
can eliminate the possibility of birefringence in the Casimir vacuum
and place strong constraints on the general features of photon
propagation in this vacuum. Our results are in fact generic to any
form of nonlinear electrodynamics subject to the boundary conditions
appropriate for the Casimir vacuum, and are not specifically
restricted to QED. Where QED is important is in determining the
specific form of the functions $d_1(z)$ and $d_2(z)$, which determine
$\xi(z)$, and hence determine the {\emph{size\/}} (but not the
qualitative features) of the Scharnhorst effect.

Unless one wants to perform calculations to orders higher than
$\alpha^2$, it is sufficient to consider the Euler--Heisenberg
Lagrangian~\cite{Euler-Heisenberg} which, in the ${\cal
F}$--${\cal G}$ formalism adopted above, takes the form
\begin{equation}
\label{E:ehxy} \L_{\rm EH}=-\frac{1}{4\pi}\;{\cal F}+c_1\;{\cal F
}^2+c_2\;{\cal G}^2,
\end{equation}
with
\begin{equation}
\label{E:coeff} c_1=\frac{\alpha^2}{90\pi^2 m^4_\electron}, \qquad
c_2=\frac{7 \alpha^2}{360\pi^2 m^4_\electron}.
\end{equation}
The terms proportional to ${\cal F}^2$ and ${\cal G}^2$ of this
Lagrangian are quartic in the field, and describe the low-energy
limit of the box diagram in QED, when four photons couple to a
single virtual electron loop.  Thus, the Lagrangian
(\ref{E:ehxy}) is only accurate to order $\alpha^2$, and it is
meaningless to retain higher order terms within this model.

In particular, deviations from (3+1) Lorentz symmetry due to the
presence of the plates (the plates reduce the symmetry group to
that of (2+1) Lorentz invariance) must on physical grounds vanish
as the plate separation goes to infinity, where one must recover
the full (3+1) Lorentz symmetry. On dimensional grounds this
implies that such terms will be suppressed by some function of
$\hbar/(m_e c a)$ and can, to lowest nontrivial order, be
neglected in the effective Lagrangian. On the other hand, even
for lowest nontrivial order in the effective Lagrangian (that is,
for the Euler--Heisenberg Lagrangian), the presence of the plates
leads to a nontrivial expectation value for $\langle C|
F_{\mu\nu}\;F_{\alpha\beta} |C \rangle$ and will in this way
contribute to the effective metric.

For the Euler--Heisenberg Lagrangian, the tensor
$\Omega^{\mu\nu}{}_{\alpha\beta}$ is
\begin{eqnarray}
\label{E:MEH} \Omega^{\mu\nu}{}_{\alpha\beta}&=&
\left(-{1\over16\pi}  + {c_1 \;{\cal F}\over 2} \right) \; \big(
\delta^\mu{}_\alpha\; \delta^\nu{}_\beta - \delta^\mu{}_\beta \;
\delta^\nu{}_\alpha \big)
\nonumber\\
&& + {c_2 \; {\cal G}\over2} \; \epsilon^{\mu\nu}{}_{\alpha\beta}
\nonumber\\
&&
+{c_1\over2} \; F^{\mu\nu}\;F_{\alpha\beta}
+{c_2\over2} \; \*F^{\mu\nu} \;\*F_{\alpha\beta},
\end{eqnarray}
so in the Casimir vacuum
\begin{equation}
d_1(z) = -\frac{1}{16\pi}+ O(\alpha^2)
\end{equation}
while $d_2(z)$ is of order $\alpha^2$.  Hence, to first order in
$\alpha^2$, we have $\xi(z)=-16\pi\,d_2(z)$, and
\begin{equation}
\label{E:speed2'} v_\phase(\theta) = v_{\mathrm{group}}(\theta)
=1-8\pi d_2(z) \; \cos^{2}\theta.
\end{equation}
Though in principle the coefficient $d_2(z)$ could depend on $z$,
the position relative to the two plates, we shall see that in the
specific case of the Casimir vacuum it is simply a
position-independent number.

To establish this, start with (\ref{E:porc2}), and insert the
specific form (\ref{E:MEH}).  Using the algebraic
identity~\cite{Schwinger,Dittrich-Gies}
\begin{equation}
\*F^{\mu\nu}\; \*F_{\alpha\nu} = F^{\mu\nu}\; F_{\alpha\nu}
-2\,{\cal F}\,\delta^\mu{}_\alpha,
\end{equation}
one easily gets
\begin{eqnarray}
\Omega^{\mu\nu}{}_{\alpha\nu} &=& \left(-{1\over16\pi}   + {c_1
\;{\cal F}\over 2} \right) \; 3 \; \delta^\mu{}_\alpha
\nonumber\\
&& +2\pi\left(c_1+c_2\right)T^\mu{}_\alpha
+{\left(c_1-c_2\right){\cal F}\over2}\,\delta^\mu{}_\alpha,
\end{eqnarray}
where $T^\mu{}_\alpha$ is Maxwell's stress-energy-momentum tensor.
Performing the indicated traces, and using the fact that the
Maxwell stress-energy tensor is traceless, we find
\begin{equation}
d_2(z) = \frac{4\pi}{3} (c_1+c_2) \; \left\langle C\left| T_{zz}
\right| C\right\rangle.
\end{equation}
The symmetries of the Casimir vacuum stress-energy (as analyzed
for instance by DeWitt~\cite{DeWitt}), then imply
\begin{equation}
d_2(z) = 4\pi (c_1+c_2) \; \left\langle C\left| T_{00} \right|
C\right\rangle.
\end{equation}
Finally the well-known result $\langle
C|T_{00}|C\rangle=-\pi^2/720 a^4$ allows us to write
\begin{equation}
d_2 =-\frac{11 \pi\alpha^2}{64800\, a^4\,m^4_\electron}.
\end{equation}
And in particular, this implies $d_2(z)$ is position independent.
Thus, at first order in $\alpha^2$,
\begin{equation}
\label{E:speed2} v_\phase(\theta)  = v_{\mathrm{group}}(\theta) =
1+\frac{11\pi^2\alpha^2}{8100\, a^4\,
m^4_\electron}\,\cos^{2}\theta.
\end{equation}
This expression reproduces Scharnhorst's result in the case
$\theta=0$, generalizing it to an arbitrary direction of
propagation. It is interesting to notice that the correction is
essentially determined by the expectation value of the energy
density, $\langle C|T_{00}|C\rangle$, in agreement with the
general results of~\cite{Latorre,Dittrich-Gies}.

\section{Conclusions}
We have outlined a general scheme that allows one to write down
the dispersion relation, and to find the polarization states, for
electromagnetic radiation propagating in a region where nonlinear
effects cannot be ignored.  The deviations from the behaviour in
Maxwell's theory are completely described by the tensorial
quantity $\Delta^{\mu\alpha\nu\beta}$, that has its origin in the
anisotropy of the medium.  In particular, we have investigated
the case of propagation in the Casimir vacuum, where there is a
privileged direction in space, identified by a unit vector $n$.
We have seen that symmetry considerations alone imply that
$\Delta^{\mu\alpha}{}_{\nu\beta}$ must be of the form
(\ref{E:MEHsym}).  This, in turn, implies that there is only one
value for the speed of light, independent of the polarization.
Thus, the possibility of birefringence in the Casimir vacuum is
completely ruled out by very general arguments (the abstract form
of the Lagrangian for nonlinear electrodynamics, plus the tensor
structure of $\Delta^{\mu\alpha}{}_{\nu\beta}$ dictated by the
geometry). Because of its generality, this conclusion applies not
just to QED itself but to arbitrary types of nonlinear
electrodynamics (such as for instance, Born--Infeld theories).

How are we to interpret this result in view of the fact that
nonlinear electrodynamics generically {\em does\/} lead to
birefringence~\cite{Adler,Novello,Novello-bis}? The key point is
that in those analyses the background field
$(F^{\mu\nu})_\background$ is nonzero. More generally, even after
averaging over quantum fluctuations of the background,
$\langle\psi|F^{\mu\nu}|\psi\rangle \neq 0$ in the quantum state
appropriate to those analyses.  In the Casimir vacuum on the
other hand, the expectation values linear in the field do vanish,
$\langle C|F^{\mu\nu}|C \rangle = 0$, and it is only the
quadratic expectation values (and higher) that are nonzero
($\langle C|\Omega^{\mu\nu\alpha\beta}|C\rangle\neq0$).  This key
difference makes the analyses of~\cite{Adler,Novello,Novello-bis}
inapplicable to the Casimir vacuum---technically speaking, the
polarization basis used in those papers fails to be meaningful
for the Casimir vacuum.  On the other hand, the discussion of the
present paper could be easily adapted to treat light propagation
in the presence of an electromagnetic background field. It is
sufficient to keep in mind that in this case there is a preferred
2-dimensional plane defined by the 2-form
$\langle\psi|F^{\mu\nu}|\psi\rangle$. Correspondingly, the form of
$\Delta^{\mu\alpha\nu\beta}$ will be more complicated [see
(\ref{E:M})].

We reiterate that the absence of birefringence is crucial for the
use of the ``effective geometry'' approach (adopted in this type
of context by Latorre {\emph{et al\/}}~\cite{Latorre}, and
further developed by Novello {\emph{et al\/}}~\cite{Novello}, and
by Dittrich and Gies~\cite{Dittrich-Gies}). Only if the
propagation of light does not depend on its polarization and is
thus, in a sense, universal, it is meaningful to describe it by a
single effective metric.  As a consequence of our analysis,
photon propagation in the Casimir vacuum can indeed be phrased
entirely in terms of the effective metric (\ref{E:metrica}). This
observation is potentially important in that ``effective metric''
approaches similar in spirit to the above are currently
attracting attention in fields as diverse as
acoustics~\cite{Unruh,Acoustics}, optics~\cite{Leonhardt,Optics},
superfluid quasiparticles~\cite{Volovik}, and Bose--Einstein
condensates~\cite{Garay}.

Finally, we mention related work on the light cone condition for
a thermalized QED vacuum due to Gies~\cite{Gies} (wherein the
analysis implicitly relies on taking the quantum expectation
value of $\Omega^{\mu\nu\alpha\beta}$ in a manner somewhat
analogous to the present paper), and intriguing results on photon
propagation in rather general linear theories in classical
backgrounds due to Obukhov and
co-workers~\cite{Obukhov-Hehl,Hehl,Obukhov}.

\section*{Acknowledgements}

It is a pleasure to thank Holger Gies and Klaus Scharnhorst for
several extremely useful comments that led to improvements in the
presentation.  MV was supported by the US Department of Energy.

\appendix

\section{Connection with other formulations}
\label{app:schwinger}

We have tried to set up the formalism in a streamlined and
self-contained manner. Nevertheless to aid comparison with other
results in the literature it is useful to give the explicit form
of the tensor $\Omega^{\mu\nu\alpha\beta}$ for generic
Schwinger-like Lagrangians of the form
(\ref{E:Schwinger-lagrangian}). For such Lagrangians
\begin{eqnarray}
\label{E:Schwinger-like-omega} \Omega^{\mu\nu\alpha\beta} &=&
{1\over4}\;\left( \partial_{\cal F}\L \right) \; \big(
\eta^{\mu\alpha}\; \eta^{\nu_\beta} - \eta^{\mu\beta} \;
\eta^{\nu\alpha} \big)
\nonumber\\
&&+{1\over4}\;\left( \partial_{\cal G}\L \right)
\;\epsilon^{\mu\nu\alpha\beta}
+{1\over4}\;M^{\mu\nu\alpha\beta},
\end{eqnarray}
where
\begin{eqnarray}
\label{E:M} M^{\mu\nu\alpha\beta}&=& F^{\mu\nu}\;F^{\alpha\beta}\;
\left(\partial^{2}_{\cal F}\L\right)+
\*F^{\mu\nu}\;\*F^{\alpha\beta}\;\left(\partial^2_{\cal G}
\L\right)
\nonumber\\
&& +\left(F^{\mu\nu}\;\*F^{\alpha\beta}+
\*F^{\mu\nu}\;F^{\alpha\beta}\right)\partial_{\cal FG}\L
\end{eqnarray}
has the same symmetries as $\Omega^{\mu\nu\alpha\beta}$.

As soon as one inserts this tensor into the photon equation of
motion (\ref{E:eom3}), the completely antisymmetric part
proportional to the Levi--Civita tensor drops out, because of the
Bianchi identity (\ref{E:bianchi2}). The remaining pieces
reproduce the photon equation of motion in the perhaps more usual
form considered by Dittrich and Gies~\cite{Dittrich-Gies}, or
Novello and co-workers~\cite{Novello,Novello-bis}.

Also note that, depending on the details of the geometry and the
quantum state, $M^{\mu\nu\alpha\beta}$ can contribute to both
$d_1$ and to $\Delta^{\mu\nu\alpha\beta}$.

\section{Fresnel equation in nonlinear electrodynamics}
\label{app:fresnel}

Our purpose is to prove equation (\ref{eq:poli}). The spatial
components of the matrix (\ref{E:matrix}) are
\begin{eqnarray}
A^{ij} &=& \omega^2 \; \langle\psi|\Omega^{i0j0}|\psi\rangle -
\omega\; k_m \; \langle\psi|\Omega^{i0jm} +
\Omega^{imj0}|\psi\rangle\nonumber\\
&&\qquad + k_m k_n \; \langle\psi|\Omega^{imjn}|\psi\rangle.
\end{eqnarray}
Let us define the unit vector $\hat{k}=\vec{k}/|\vec{k}|$. Then
the components of $A^{ij}$ in a basis with one axis directed along
$\hat{k}$ are
\begin{equation}
A^{ij} \hat{k}_j = \omega^2 \;
\langle\psi|\Omega^{i0j0}|\psi\rangle \; \hat{k}_j - \omega\; k_m
\; \langle\psi|\Omega^{imj0}|\psi\rangle \; \hat{k}_j \equiv
\omega \; V^{i}.
\end{equation}
In particular
\begin{equation}
A^{ij} \hat{k}_i \hat{k}_j = \omega^2 \;
\langle\psi|\Omega^{i0j0}|\psi\rangle \; \hat{k}_i \hat{k}_j
\equiv \omega^2 \; S.
\end{equation}
Then the matrix $A^{ij}$ has the following structure:
\begin{equation}
  \label{eq:matr}
  \left(
  \begin{array}{cc}
    \omega^2 \; S & \omega \; V^{J}\\
    \omega \; V^{I} & T^{IJ}
  \end{array}
         \right),
\end{equation}
where $I$ and $J$ label the two directions orthogonal to
$\hat{k}$. Evaluating the determinant by expanding in the first
row or column, it is easy to see that every term will contain at
least two factors of $\omega$, which establishes equation
(\ref{eq:poli}) as desired.

\section{Polarization sum}
\label{app:polsum}

In this appendix we discuss the reason for the difference between
our expressions (\ref{E:metrica}) and (\ref{E:nuova!}) for the
effective metric and those implicit in the ``light cone
condition'' in reference~\cite{Dittrich-Gies}.  In the spirit of
that paper a dispersion relation would be derived as follows (a
certain amount of ``translation'' is required to get that
formalism to match with the current notation).  First, one writes
equation (\ref{E:eom-master2}) for two polarization states
$\widetilde{\epsilon}^{\,(1)}$ and $\widetilde{\epsilon}^{\,(2)}$
orthogonal to each other:
\begin{equation}
\label{E:?} \langle\psi|\Omega^{\mu\alpha\nu\beta}|\psi\rangle\,\;
k_\alpha\,k_\beta\,\widetilde{\epsilon}^{\,(r)}_{\nu}=0.
\end{equation}
Working in Lorentz gauge $k\cdot\widetilde{\epsilon}^{\,(r)}=0$
this can be written as
\begin{equation}
\label{E:?2} d_1 \; k^2 \; \widetilde{\epsilon}_{(r)}^{\;\mu} +
\Delta^{\mu\alpha}{}_{\nu\beta}\;
k_\alpha\,k^\beta\,\widetilde{\epsilon}_{(r)}^{\;\nu}=0,
\end{equation}
where $\widetilde{\epsilon}_{(r)}^{\;\mu}=
\eta^{\mu\nu}\,\widetilde{\epsilon}^{\,(r)}_{\nu}$. One then
takes the scalar product of each equation with the corresponding
polarization vector. Finally, the resulting two equations are
summed. Furthermore, it was explicitly asserted that one could
effectively replace~\cite{Dittrich-Gies}
\begin{equation}
\label{E:wrong-sum} \sum_{r=1}^2
\widetilde{\epsilon}_{\,(r)}^{\;\mu}\;
\widetilde{\epsilon}_{(r)}^{\;\nu} \to \eta^{\mu\nu}
\end{equation}
in the terms containing $\Delta^{\mu\alpha}{}_{\nu\beta}$. Under this
hypothesis one obtains, following the steps described above,
\begin{equation}
\label{E:DG} 2 d_1 k^2 +
\Delta^{\mu\alpha}{}_{\nu\alpha}\,
k_\mu\,k^\nu=0,
\end{equation}
which is equivalent to equation (14) of
reference~\cite{Dittrich-Gies}. This has the form of the
dispersion relation $\widetilde{\gamma}^{\,\mu\nu}\;k_\mu
\,k_\nu=0$, with
\begin{equation}
\label{E:DG2} \widetilde{\gamma}^{\,\mu\nu} \propto 2\,d_1\,
\eta^{\mu\nu} + \Delta^{\mu\alpha\nu\beta}\, \eta_{\alpha\beta}.
\end{equation}
Specializing to the Casimir vacuum, and normalizing appropriately, one
has
\begin{equation}
\label{E:wrongmetric} \widetilde{\gamma}^{\,\mu\nu}=\eta^{\mu\nu}+
\frac{2\,d_2}{2\,d_1+d_2}\,n^\mu\,n^\nu,
\end{equation}
clearly different from (\ref{E:metrica}). Nevertheless, when one
considers QED corrections to the Maxwell Lagrangian, $d_1$ is of order
$\alpha^0$, while $d_2$ is of order $\alpha^2$ (see
section~\ref{S:size} above); therefore, the two metrics
(\ref{E:metrica}) and (\ref{E:wrongmetric}) agree to order $\alpha^2$.

The reason for the discrepancy is that the formal replacement
(\ref{E:wrong-sum}) is limited in a subtle manner.  Strictly
speaking, it is correct as it stands only if photon propagation
is governed by the Minkowski metric $\eta_{\mu\nu}$; which is
exactly the situation we are trying to get away from. In order to
see this explicitly, let us introduce an orthonormal tetrad
$\left(u, \widetilde{\epsilon}_{(1)}, \widetilde{\epsilon}_{(2)},
w\right)$, with $u$ timelike and $w$ spacelike, so one can write
\begin{equation}
\label{E:33} \sum_{r=1}^2 \widetilde{\epsilon}_{(r)}^{\;\mu}\;
\widetilde{\epsilon}_{(r)}^{\;\nu} = \eta^{\mu\nu} + u^\mu\,u^\nu
- w^\mu\,w^\nu .
\end{equation}
Working in the Lorentz gauge, as in
reference~\cite{Dittrich-Gies}, the two-dimensional subspace of
Minkowski spacetime, orthogonal to $\widetilde{\epsilon}_{(1)}$
and $\widetilde{\epsilon}_{(2)}$ and spanned by the unit vectors
$u$ and $w$, contains the wave vector $k$, so
\begin{equation}
\label{E:k} k^\mu=-\left(k\cdot u\right)\,u^\mu+\left(k\cdot
w\right)\,w^\mu.
\end{equation}
Following the same steps that led to (\ref{E:DG}), but using
(\ref{E:33}) instead of the replacement (\ref{E:wrong-sum}), we find,
in place of the single term $\Delta^{\mu\alpha}{}_{\nu\alpha}\,
k_\mu\,k^\nu$ occurring in (\ref{E:DG}), the two terms
\begin{equation}
\label{E:omegakkright}
\Delta^{\mu\alpha}{}_{\nu\alpha}\,k_\mu\,k^\nu
+k^2\;\Delta_{\mu\alpha\nu\beta}\,
u^\mu\,u^\nu\,w^\alpha\,w^\beta,
\end{equation}
where we have used (\ref{E:k}) and the symmetry properties of
$\Delta_{\mu\alpha\nu\beta}$.  Clearly, the second term in the
expression (\ref{E:omegakkright}) cannot be ignored unless
$k^2=0$. We conclude that it is the subtly incorrect replacement
(\ref{E:wrong-sum}) which implies that the derivation of
(\ref{E:DG}), (\ref{E:DG2}), and (\ref{E:wrongmetric}) is limited
to first order in $\alpha^2$.  For the Casimir vacuum,
\begin{eqnarray}
\label{E:add} \Delta_{\mu\alpha\nu\beta}\;
u^\mu\,u^\nu\,w^\alpha\,w^\beta
&=&
d_2 \left[ (n\cdot u)^2 - (n\cdot w)^2 \right]
\nonumber\\
&=& d_2\left[ -1 +
\sum_{r=1}^2\left(n\cdot\widetilde{\epsilon}_{(r)}\right)^2
\right].
\nonumber\\
&&
\end{eqnarray}
Using now the relation
$k^2\left(n\cdot\widetilde{\epsilon}_{(r)}\right)=0$, which
follows from (\ref{E:?2}) upon contraction with $n$ in the
Lorentz gauge, we see that
\begin{equation}
k^2 \; \Delta_{\mu\alpha\nu\beta}u^\mu u^\nu w^\alpha w^\beta
=-d_2 \; k^2,
\end{equation}
so (\ref{E:DG}), corrected by the additional term (\ref{E:add}),
indeed reproduces the metric (\ref{E:metrica}).

In general, if the photon dispersion is in fact determined by some
(unique) effective metric ${\rm g}_{\mu\nu}$ then one can choose a
tetrad like $\left(u, \widetilde{\epsilon}_{(1)},
\widetilde{\epsilon}_{(2)}, w\right)$, but which is orthonormal
with respect to ${\rm g}_{\mu\nu}$, so now
\begin{equation}
\label{E:33'} \sum_{r=1}^2 \widetilde{\epsilon}_{(r)}^{\;\mu}\;
\widetilde{\epsilon}_{(r)}^{\;\nu} = \gamma^{\,\mu\nu} +
u^\mu\,u^\nu - w^\mu\,w^\nu .
\end{equation}
Applying this polarization sum to the photon propagation equation
written in the form (\ref{E:eom-master2}), and using the properties
(\ref{E:eomfc}) and (\ref{E:gkepsilon}), we see that the strictly
correct replacement for the polarization sum is
\begin{equation}
\label{E:correct-sum} \sum_{r=1}^2
\widetilde{\epsilon}_{(r)}^{\;\mu}\;
\widetilde{\epsilon}_{(r)}^{\;\nu} \to \gamma^{\,\mu\nu}.
\end{equation}
This modified polarization sum, applied to (\ref{E:eom-master2}), is
consistent with the ``bootstrap'' condition (\ref{E:bootstrap2}).

In the particular case of the Casimir vacuum, it is a simple
matter to verify, using our expression for $\gamma^{\,\mu\nu}$,
that the bootstrap condition is indeed satisfied. It is now also
clear why the formalism of~\cite{Dittrich-Gies} works to first
order in $\alpha^2$: Since the difference between $\eta^{\mu\nu}$
and $\gamma^{\,\mu\nu}$ is itself of order $\xi$ (that is, of
order $\alpha^2$) the difference between the (subtly incorrect)
formula (\ref{E:wrongmetric}), derived on the basis of
(\ref{E:wrong-sum}), and the consistency condition
(\ref{E:bootstrap2}) is automatically of order $\xi^2$
($\alpha^4$). This may also be explicitly verified using the
approximation $\xi \ll 1$ so that
\begin{equation}
{\rm g}_{\mu\nu} \approx \eta_{\mu\nu} - \xi \; n_\mu \; n_\nu.
\end{equation}
Substituting this into (\ref{E:bootstrap2}) self-consistently
reproduces $\gamma^{\,\mu\nu}$ to the required order. Note that in
almost all cases of interest one is quite content to work to
first order in $\alpha^2$ in which case all of these subtleties
are moot.

We conclude with some general comments about the derivation of a
metric from a polarization sum: If there is only one dispersion
relation for all polarizations, as for instance in the Casimir
vacuum discussed above, then $\widetilde{\epsilon}^{\,(1)}$ and
$\widetilde{\epsilon}^{\,(2)}$ both satisfy the {\emph{same}}
equation
\begin{equation}
\label{E:nonsingpol}
\overline{A}^{\,\mu\nu}(k)\;\widetilde{\epsilon}^{\,(r)}_{\;\nu}=0
\end{equation}
[see (\ref{E:nonsing})], with the {\emph{same}} matrix
$\overline{A}^{\,\mu\nu}(k)$. However, if
$\widetilde{\epsilon}^{\,(1)}$ and $\widetilde{\epsilon}^{\,(2)}$
correspond to two different dispersion relations, their
propagation equations will actually differ, because now
$\overline{A}_{(1)}^{\,\mu\nu}\neq \overline{A}_{(2)}^{\,\mu\nu}$.
Therefore, a procedure of the form described at the beginning of
this appendix is fundamentally flawed in the general case, as it
fails to take into account the possibility of birefringence, and
is in fact (strictly speaking) incompatible with birefringence.
Indeed, summing the equations
\begin{equation}
\overline{A}_{(1)}^{\,\mu\nu}\;\widetilde{\epsilon}^{\,(1)}_{\mu}\;
\widetilde{\epsilon}^{\,(1)}_{\nu}=0 \qquad \hbox{and} \qquad
\overline{A}_{(2)}^{\,\mu\nu}\;\widetilde{\epsilon}^{\,(2)}_{\mu}\;
\widetilde{\epsilon}^{\,(2)}_{\nu}=0,
\end{equation}
which do take birefringence into account, appears rather
impractical, and it is not obvious whether the structure of the
resulting equation will allow the use of a polarization sum at
all. Thus, producing ``average dispersion relations'' by
polarization sums is at some deep level internally inconsistent,
except in the cases when no polarization sum is needed. It is
only if one is limiting interest to lowest-order ($\alpha^2$)
corrections away from Minkowski space that the polarization sum
makes sense for a birefringent system, and only then in the sense
that
\begin{equation}
\label{E:wrong-sum2} \sum_{r=1}^2
\widetilde{\epsilon}_{(r)}^{\;\mu}\;
\widetilde{\epsilon}_{(r)}^{\;\nu} \to \eta^{\mu\nu} +
O(\alpha^2).
\end{equation}
That is: Although equation (\ref{E:eom-master2}) is exact, we could
choose to contract it with {\emph{approximate}} polarization states
appropriate to flat Minkowski space. As long as the deviations from
the ordinary speed of light are small, so are the deviations of the
true polarization states from the usual ones, and then an approximate
polarization sum of the above form is useful. We conclude that the
polarization sum technique is most useful if there is no
birefringence, and that in the presence of birefringence it is, at
best, only useful when strictly limited to lowest-order corrections to
Minkowski space propagation.


\end{document}